\documentclass[journal,10pt]{IEEEtran}
\IEEEoverridecommandlockouts
\usepackage{cite}
\usepackage[ruled,vlined]{algorithm2e}
\usepackage{xfrac}
\usepackage{amsmath,amssymb,amsfonts}
\usepackage{algorithmic}
\usepackage{graphicx}
\usepackage{textcomp}
\usepackage[svgnames]{xcolor}
\usepackage{enumitem}
\usepackage{cancel}
\usepackage{subfig}
\usepackage[nolist,nohyperlinks]{acronym}

\ifCLASSINFOpdf
 
\else
 
\fi
\def\gil#1{{\color{black}#1}}
\def\gild#1{{\color{black}#1}}

\def\gilm#1{{\color{black}#1}}
\newcommand{\gf}[1]{\textcolor{black}{{#1}}}
\begin{document}
\begin{acronym}
  \acro{IRS}{Intelligent reflecting surface}
  \acro{RIS}{reconfigurable intelligent surface}
   \acro{irs}{intelligent reflecting surface}
  \acro{PARAFAC}{parallel factor}
  \acro{TALS}{trilinear alternating least squares}
    \acro{BALS}{bilinear alternating least squares}
  \acro{DF}{decode-and-forward}
  \acro{AF}{amplify-and-forward}
  \acro{CE}{channel estimation}
  \acro{RF}{radio-frequency}
  \acro{THz}{Terahertz communication}
  \acro{EVD}{eigenvalue decomposition}
  \acro{CRB}{Cramér-Rao lower bound}
  \acro{CSI}{channel state information}
  \acro{BS}{base station}
  \acro{MIMO}{multiple-input multiple-output}
   \acro{NMSE}{normalized mean squared error}
   \acro{2G}{Second Generation}
  \acro{3G}{3$^\text{rd}$~Generation}
  \acro{3GPP}{3$^\text{rd}$~Generation Partnership Project}
  \acro{4G}{4$^\text{th}$~Generation}
  \acro{5G}{5$^\text{th}$~Generation}
  \acro{6G}{6$^\text{th}$~generation}
  \acro{E-TALS}{\textit{enhanced} TALS}
  \acro{UT}{user terminal}
  \acro{UTs}{users terminal}
  \acro{LS}{least squares}
  \acro{KRF}{Khatri-Rao factorization}
  \acro{KF}{Kronecker factorization}
  \acro{MU-MIMO}{multi-user multiple-input multiple-output}
  \acro{MU-MISO}{multi-user multiple-input single-output}
 \acro{MU}{multi-user}
 \acro{SER}{symbol error rate}
 \acro{SNR}{signal-to-noise ratio}
 \acro{SVD}{singular value decomposition}
\end{acronym}
\title{Channel Estimation for Intelligent Reflecting Surface Based MIMO Communication Systems via Tensor Modeling}
\title{Channel Estimation for Intelligent Reflecting Surface Based MIMO Communication Systems: A Tensor Modeling Approach}
\title{Tensor Modeling Approach to Receiver Design for Intelligent Reflecting Surface Assisted MIMO Systems}
\title{A Tensor Approach to Intelligent Reflecting Surface Assisted MIMO Systems: Semi-Blind Joint Channel and Symbol Estimation }
\title{Semi-Blind Joint Channel and Symbol Estimation in IRS-Assisted MIMO Systems}
\title{Semi-Blind Joint Channel and Symbol Estimation in Intelligent Surface Assisted MIMO Systems}
\title{Semi-Blind Joint Channel and Symbol Estimation in IRS-Assisted Multi-User MIMO Networks}

\author{Gilderlan~T.~de~Ara\'{u}jo, Paulo~R.~B.~Gomes,
      Andr\'{e}~L.~F.~de~Almeida,~\IEEEmembership{Senior~Member,~IEEE}, Gábor~Fodor,~\IEEEmembership{Senior~Member,~IEEE} and~Behrooz Makki,~\IEEEmembership{Senior~Member,~IEEE}
 \thanks{Gilderlan T. de Ara\'{u}jo, Paulo R. B. Gomes and Andr\'{e} L. F. de Almeida are with Wireless Telecommunication Research Group (GTEL), Department of Teleinformatics, Federal University of Cear\'{a}, Fortaleza, CE, e-mails: \{gilderlan,paulo,andre\}@gtel.ufc.br.}
 \thanks{Gábor Fodor is with Ericsson Research, 16480 Stockholm, Sweden, and	also with the Division of Decision and Control, KTH Royal Institute of 	Technology, 11428 Stockholm, Sweden, e-mail: gabor.fodor@ericsson.com.} 
 \thanks{Behrooz Makki is with Ericsson Research, Ericsson, 417 56 Göteborg, Sweden, e-mail: behrooz.makki@ericsson.com.}
 \thanks{This work was supported in part by the Ericsson Research, Technical Cooperation UFC.48 and in part by the Coordenação de Aperfeiçoamento de Pessoal de Nível Superior - Brasil (CAPES)-Finance Code 001, and CAPES/PRINT Proc. 88887.311965/2018-00. Andr\'{e}~L.~F.~de~Almeida acknowledge CNPq for its financial support under the grant 312491/2020-4.}
}


\markboth{Journal of \LaTeX\ Class Files,~Vol.XX, No.~X, XXX}%
{Shell \MakeLowercase{\textit{et al.}}: Bare Demo of IEEEtran.cls for IEEE Journals}

\maketitle

\begin{abstract}
\ac{IRS} is a promising technology for beyond \acl{5G} of the wireless communications. In fully passive \ac{IRS}-assisted systems, \acl{CE} is challenging and should be carried out only at the base station or at the terminals since the elements of the \ac{IRS} are incapable of processing signals. In this letter, we formulate a tensor-based semi-blind receiver that solves the joint channel and symbol estimation problem in an \ac{IRS}-assisted \acl{MU-MIMO} system. The proposed approach relies on a generalized PARATUCK tensor model of the signals reflected by the \ac{IRS}, based on a two-stage closed-form semi-blind receiver using Khatri-Rao  and Kronecker factorizations. Simulation results demonstrate the superior performance of the proposed semi-blind receiver, in terms of the normalized mean squared error and \acl{SER}, as well as a lower computational complexity, compared to recently proposed \acl{PARAFAC} analysis-based receivers.
\end{abstract}


\IEEEpeerreviewmaketitle

\section{Introduction}
\label{introduction}
\IEEEPARstart{I}{ntelligent} reflecting surface is a promising technology to beyond \ac{5G} of the  wireless networks, which may offer high spectral efficiency, while improving the energy efficiency/reliability and reducing the end-to-end latency and cost \cite{Behrooz2022}. An \ac{IRS} is a two-dimensional array structure composed of a large number of passive (or semi-passive) software-controlled reconfigurable scattering elements, whose electromagnetic response can be dynamically adjusted \cite{Basar2021}. In most studied implementations, an \ac{IRS} operates by applying phase shifts to the incident radio waves in favor of signal reception. 

In an \ac{IRS}-assisted wireless network, the acquisition of instantaneous \ac{CSI} is an important and challenging task, since the accuracy of the \ac{CSI} has significant impact on the optimization of the \ac{IRS} phase shifts. Recent works have proposed different solutions to tackle the \ac{CE} problem in \ac{IRS}-assisted communications in single-user single-antenna/multi-antenna systems \cite{Gil_JTSP,Chen2021}. \gild{In \cite{Gil_JTSP}, the \ac{CE} performance \gf{is evaluated} by means of a tensor modeling of the received signal using an iterative solution, which increases the complexity. In \cite{Chen2021}, the authors \gf{assume} a semi-active \ac{IRS} structure. This assumption undermines the low-cost \gf{structure} of the \ac{IRS}, since \ac{RF} chains \gf{are} used.} Considering a \ac{MU} scenario, \cite{Zheng2020} and \cite{Wang2020} exploit \ac{IRS} element grouping and spatial correlation to control the pilot overhead. \gild{In both \cite{Zheng2020} and \cite{Wang2020}, the authors assume a quasi-static block fading channel model for all the involved links, which leads to a degradation on the performance of \ac{CE} when some mobility of the \ac{UT} is \gf{considered}.} In particular, \cite{Wang2020} \gf{exploits} the correlation among the \ac{UT}-\ac{IRS}-\ac{BS} \gf{channels of different users}. \gf{Note that all these methods are pilot-assisted schemes.} \gf{In contrast,} blind and semi-blind receivers perform \ac{CE} and data detection without \gf{employing} pilot sequences.

Approaches based on tensor modeling have been proposed for conventional point-to-point \ac{MIMO} systems (see\gil{,} e.g.\gil{,} \cite{Almeida2008,Favier2014} and the references therein). \gf{Specifically}, a constrained factor decomposition is derived in \cite{Almeida2008} to formulate a space-time spreading model, while \cite{Favier2014} capitalizes on the PARATUCK\footnote{\gild{The name PARATUCK is derived from the combination of the \acs{PARAFAC} and Tucker tensor decompositions.}} model to derive a semi-blind joint \ac{CE} and data detection for multi-carrier \ac{MIMO} systems. 
In the context of \ac{IRS}-assisted communications, tensor modeling has been recently proposed in \cite{Alexander_SAM2020} to solve the \ac{CE} problem in a \ac{MU-MISO} \ac{IRS}-assisted system, by capitalizing on the multidimensional structure of the signal reflected by the \ac{IRS} via a pilot-assisted scheme based on the \ac{PARAFAC} tensor decomposition. \gf{Reference} \cite{Gil_JTSP} goes in the same direction, by proposing iterative and closed-form \ac{CE} methods for single-user \ac{MIMO} \ac{IRS}-assisted communications. These works, however, can only operate with the use of pilot sequences, at the cost of increasing the end-to-end latency.

This letter\footnote{\textit{Notation}: Scalars are denoted by lowercase letters ($a$), vectors by bold lowercase letters ($\mathbf{a})$, matrices by bold capital letters ($\mathbf{A})$, and tensors by calligraphic letters $(\mathcal{A})$. The transpose and Hermitian transpose of a given matrix $\mathbf{A}$ are denoted by $\mathbf{A}^{\textrm{T}}$ and $\mathbf{A}^{\textrm{H}}$, respectively. $D_i(\mathbf{A})$ is a diagonal matrix holding the $i$-th row of $\mathbf{A}$ on its main diagonal. $\| \cdot \|_{\text{F}}$ is the Frobenius norm of a matrix. The operator $\textrm{diag}(\mathbf{a})$ forms a diagonal matrix out of its vector argument, while $\diamond$ and $\otimes$ denote the  Khatri Rao and Kronecker products, respectively. The operator $\textrm{vec}(\cdot)$ vectorizes an $I \times J$ matrix argument, while $\textrm{unvec}_{I \times J}(\cdot)$ does the opposite operation. $\mathbf{A}_{i.}$ denotes the $i$-th row of the matrix $\mathbf{A}$, and $\mathbf{I}_{J}$ is an identity matrix of size $J \times J$.} formulates a semi-blind receiver that solves the problem of joint channel and symbol estimation in an \ac{IRS}-assisted wireless network without the need of \gild{a dedicated training stage}. We address a realistic scenario, in which the channels between the \ac{UT} and the \ac{IRS} undergo shorter term variations compared to the channel between the \ac{BS} and the \ac{IRS}. Considering an uplink \ac{MU-MIMO} setup, we show that the signals reflected by the \ac{IRS} and received at the \ac{BS} follow a generalized PARATUCK tensor model. By exploiting the algebraic structure of this tensor model, we derive a semi-blind receiver algorithm that allows the \ac{BS} to jointly estimate the multiple uplink \ac{UT}-\ac{IRS} channels, the common \ac{IRS}-\ac{BS} channel, and the data symbols transmitted by all \ac{UT}s, in a closed-form way by solving simple rank-one matrix approximation problems. 

The proposed two-stage closed-form semi-blind receiver consists of a sequential combination of the \ac{KRF} and \ac{KF} schemes, and is referred to as KAKF \gf{in the sequel}. 
We compare our proposed semi-blind method with \gf{a recently} proposed \ac{PARAFAC}-based \gf{receiver} \gild{\cite{Alexander_SAM2020} and with the \ac{KRF} receiver \cite{Gil_JTSP}, which are two pilot-assisted methods. The first is an iterative solution  that solves two \ac{LS} problems to estimate the involved channel matrices, while the second is a  closed-form scheme based on rank-one matrix approximations.} Numerical results corroborate the improved \ac{CE} accuracy, superior \ac{SER} performance, and lower computational complexity offered by the proposed semi-blind receiver.

\section{System Model}
\label{Sec: System_Model}

\begin{figure}[!t]
    \centering
    \includegraphics[scale = 0.4]{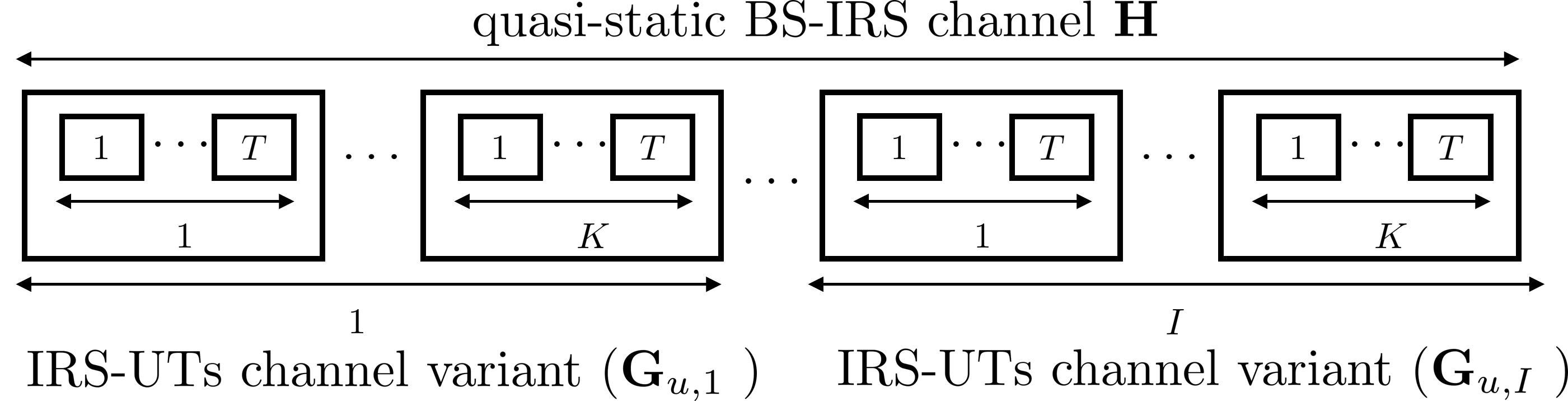}
    \caption{Transmission structure.}
    \label{fig:time protocol}
\end{figure}

Consider the uplink communication in an \ac{IRS}-assisted \ac{MU-MIMO} system, in which the \ac{BS} is equipped with $M$ antennas, the \ac{IRS} is composed of $N$ passive scattering elements, while each of the $U$ \ac{UT}s has $L$ antennas\footnote{We assume that the users have the same number of antennas for simplicity of exposition. However, the proposed solution can be straightforwardly adapted to users with different numbers of antennas.}. We assume that the \ac{BS} and the \ac{IRS} are deployed at  fixed heights, e.g., on the roof of a building, compared to the moving \ac{UT}s.  In this sense, we assume that  the transmitted data is organized into $I$ data frames. Each frame is composed of $KT$ symbol periods, where $K$ denotes the number of blocks contained in each frame and $T$ is the number of time slots in each block. This transmission structure is illustrated in  Fig. \ref{fig:time protocol}. 
To capture a certain level of \ac{UT} mobility, we assume that the \ac{UT}-\ac{IRS} channels stay constant during a frame, but vary in different frames independently. On the other hand, the \ac{IRS}-\ac{BS} channel follows a quasi-static model \gild{\cite{Hu2021}}, and is assumed to remain constant during the transmission time. This is a reasonable assumption since the \ac{BS} and the \ac{IRS} are assumed to be deployed at fixed positions. We also assume that the direct communication links between the \ac{UT}s and the \ac{BS} are weak or not available. 
Each \ac{UT} sends $L$ data symbol vector that are \textit{diagonally} coded as $\mathbf{z}_u[k,t] = \textrm{diag}(\mathbf{w}_u[k])\mathbf{x}_u[t] \in \mathbb{C}^{L \times 1}$, where $\mathbf{x}_u[t] \in \mathbb{C}^{L \times 1}$ is the symbol vector transmitted by the $u$-th \ac{UT} at the $t$-th time slot, $t = 1, \dots, T$, whose entries are drawn from any finite-alphabet constellation. The vector $\mathbf{w}_u[k] \in  \mathbb{C}^{L \times 1}$ denotes the coding vector associated with the $u$-th \ac{UT} at the $k$-th block. The quasi-static \ac{IRS}-\ac{BS} channel matrix is denoted by $\mathbf{H} \in \mathbb{C}^{M \times N}$, while the time-varying channel matrix between the \ac{IRS} and the $u$-th \ac{UT} at  frame $i$ is denoted as $\mathbf{G}_{u,i} \in \mathbb{C}^{N \times L}$. Therefore, the received signal at the \ac{BS} coming from the $u$-th \ac{UT} via \ac{IRS} is given by
\begin{equation}
  \mathbf{y}_u[i,k,t] =  \mathbf{H}\textrm{diag}(\mathbf{s}[k])\mathbf{G}_{u,i}\mathbf{z}_u[k,t] + \mathbf{v}[i,k,t],
   \label{Eq:Receiver_Signal from u-th User}
\end{equation}
where $\mathbf{s}[k] \in \mathbb{C}^{N \times 1}$ collects the phase shifts $s_n[k]=e^{j\phi_n[k]}$ applied by the \ac{IRS} during the $k$-th block, and $\mathbf{v}[i,k,t]  \in \mathbb{C}^{M \times 1}$ denotes the additive white Gaussian noise (AWGN) term. Note that the \ac{IRS} phase shifts configuration and the coding vectors are known by the \ac{BS} and remains constant during the $T$ time slots within the $k$-th block, but varies between different blocks.  

At the \ac{BS}, the received signal is represented as the superposition of the signals coming from the $U$ \ac{UT}s. Therefore, the total received signal can be expressed as
\begin{equation}
\small
  \mathbf{y}[i,k,t]  = \mathbf{H}\textrm{diag}(\mathbf{s}[k])\left(\sum_{u = 1}^{U}\mathbf{G}_{u,i}\textrm{diag}(\mathbf{w}_u[k])\mathbf{x}_u[t] \right)+ \mathbf{v}[i,k,t],\\  
   \label{Eq:Receiver_Signal at t-ésimo time slot}
\end{equation}
or, more compactly,
\begin{equation}
  \mathbf{y}[i,k,t] 
  = \mathbf{H}\textrm{diag}(\mathbf{s}[k])\mathbf{G}_{i}\textrm{diag}(\mathbf{w}[k])\mathbf{x}[t] + \mathbf{v}[i,k,t], 
  \label{Eq:Receiver_Signal at t-ésimo time slot_2}
\end{equation}
where $\mathbf{G}_{i} = \left[\mathbf{G}_{1,i}, \ldots, \mathbf{G}_{U,i} \right] \in \mathbb{C}^{N \times UL}$ represents the augmented \acl{MU} uplink channel matrix  at frame $i$. Also, $\mathbf{x}[t] = \left[\mathbf{x}^{\textrm{T}}_{1}[t], \ldots, \mathbf{x}^{\textrm{T}}_{U}[t] \right]^{\textrm{T}}  \in \mathbb{C}^{UL \times 1}$ collects the transmitted data symbol vectors from all \ac{UT}s, and $\mathbf{w}[k] = \left[\mathbf{w}^{\textrm{T}}_{1}[k]\; \dots \; \mathbf{w}^{\textrm{T}}_{U}[k] \right]^{\textrm{T}}  \in \mathbb{C}^{UL \times 1}$ collects the $U$ coding vectors used by the \ac{UT}s in block $k$. Defining $\mathbf{Y}[i,k]\doteq [\mathbf{y}[i,k,1], \ldots, \mathbf{y}[i,k,T]] \in \mathbb{C}^{M \times T}$ that collects the received signal vectors during the $t = 1, \ldots, T$ time slots, we obtain
\begin{equation}
  \mathbf{Y}[i,k]  =  \mathbf{H}D_k(\mathbf{S})\mathbf{G}_{i}D_k(\mathbf{W})\mathbf{X}^{\textrm{T}} + \mathbf{V}[i,k],
   \label{Eq:Receiver_Signal at k-ésimo time slot}
\end{equation}
where $\mathbf{S} \in \mathbb{C}^{K \times N}$ is the phase shifts matrix, $\mathbf{W} = \left[\mathbf{w}[1] \, \dots \, \mathbf{w}[K]\right]^{\textrm{T}} \in \mathbb{C}^{K \times UL}$ and $\mathbf{X} = \left[\mathbf{x}[1] \, \dots \, \mathbf{x}[T]\right]^{\textrm{T}} \in \mathbb{C}^{T \times UL}$.
The useful part of the received signal in (\ref{Eq:Receiver_Signal at k-ésimo time slot}) can be identified as a generalized PARATUCK decomposition of a fourth-order tensor $\mathcal{Y} \in \mathbb{C}^{M \times I \times K \times T}$  written in terms of its $(i,k)$-th matrix slices, where the frame $i$ and block $k$ dimensions are fixed. Our goal is to jointly estimate in a semi-blind way, i.e., without resorting to a \gild{dedicated training stage,} the \ac{IRS}-\ac{BS} channel $\mathbf{H}$, the \ac{UT}s-\ac{IRS} channels that make up $\mathbf{G}_{i}$ and the transmitted data symbols $\mathbf{X}$ from the received signal tensor in (\ref{Eq:Receiver_Signal at k-ésimo time slot}). To this end, in the following\gil{,} we formulate the detailed processing stages of the proposed closed-form semi-blind receiver. 

\section{Proposed Semi-Blind KAKF Receiver}
\label{SEC: Closed-Formed Solution}
 
The vectorized form of the received signal in (\ref{Eq:Receiver_Signal at k-ésimo time slot}) as
\begin{equation}
    \begin{aligned}
        \mathbf{y}_{i,k} & = \left(\mathbf{X} \otimes \mathbf{H}\right)\textrm{vec}(D_k(\mathbf{S})\mathbf{G}_iD_k(\mathbf{W}))\\
        & = \left(\mathbf{X} \otimes \mathbf{H}\right)\left(D_k(\mathbf{W}) \otimes D_k(\mathbf{S})\right)\mathbf{g}_i
    \end{aligned}
    \label{EQ:Vec of Y_k,i},
\end{equation}
where $\mathbf{g}_i \doteq \textrm{vec}(\mathbf{G}_{i}) \,\, \in \, \mathbb{C}^{NLU}$, and we have used \textcolor{black}{the Kronecker product property} vec$(\mathbf{ABC}) = \left(\mathbf{C}^{\textrm{T}} \otimes \mathbf{A}\right)$vec$(\mathbf{B})$. Defining $\mathbf{y}_k \doteq [\mathbf{y}_{1.k}, \ldots, \mathbf{y}_{I,k}]$ that collects the $I$ received signals for the $k$-th block, we can rewrite (\ref{EQ:Vec of Y_k,i}) as
\begin{equation}
    \begin{aligned}
    \mathbf{y}_k= \mathbf{Q}\left(D_k(\mathbf{W}) \otimes D_k(\mathbf{S})\right)\mathbf{G}^{\textrm{T}}, \;{\color{black} k = 1, \dots, K,}
    \end{aligned}
    \label{EQ:y_k}
\end{equation}
where $\mathbf{G} = \left[\mathbf{g}_{1}, \ldots, \mathbf{g}_{I}\right]^{\textrm{T}} 
\in \mathbb{C}^{I \times P}$ and $\mathbf{Q} = \mathbf{X} \otimes \mathbf{H} \label{eq:defQ} \; \in \mathbb{C}^{TM \times P}$, with $P = NLU$. Resorting again to the property of the Kronecker product, we rewrite (\ref{EQ:y_k}) more compactly as 
\begin{equation}
    \mathbf{y}_k = \left(\mathbf{G} \diamond \mathbf{Q}\right)\left(\mathbf{W}_{k.} \otimes \mathbf{S}_{k.}\right)^{\textrm{T}}.
\end{equation}
Finally, by collecting the received signals $\mathbf{y}_{k}$, $k = 1, \ldots, K$, over the $K$ blocks, we have
\begin{equation}
    \mathbf{Y} \doteq [\mathbf{y}_1, \ldots, \mathbf{y}_K]= \left(\mathbf{G} \diamond \mathbf{Q}\right)\boldsymbol{\Psi} \; {\color{black}\in \mathbb{C}^{ITM \times K},} 
    \label{EQ:Y}
\end{equation}
where 
$
\boldsymbol{\Psi} = \mathbf{W}^{\textrm{T}} \diamond \mathbf{S}^{\textrm{T}} = \left[\mathbf{W}_{1.}^{\textrm{T}} \otimes \mathbf{S}_{1.}^{\textrm{T}} \, \dots \, \mathbf{W}_{K.}^{\textrm{T}} \otimes \mathbf{S}_{K.}^{\textrm{T}}\right] \in \mathbb{C}^{P \times K}.
$
From (\ref{EQ:Y}), we observe that the LS estimate of the Khatri-Rao product $\mathbf{G} \diamond \mathbf{Q}$ can be obtained\gild{. To avoid the pseudo inverse calculation,} we assume that $\boldsymbol{\Psi}$ is constructed based on a Discrete Fourier Transform factorization design\footnote{The design of the phase shifts $\mathbf{S}$ and the coding $\mathbf{W}$ matrices is a relevant point, especially in the optimization context. However, the optimization study on these matrices is out of the scope of this work. Further, the structure of $\mathbf{S}$ and $\mathbf{W}$ impacts the design of the matrix $\mathbf{\Psi}$ regarding complexity aspects.}
as in \cite{Sokal2019}, \gild{such that $\left(1/K\right)\boldsymbol{\Psi}\boldsymbol{\Psi}^{\textrm{H}} = \mathbf{I}_{P}$. Note, however, that the orthogonality of $\boldsymbol{\Psi}$ is not requirement of our semi-blind receiver.} Using $\mathbf{G} \diamond \mathbf{Q}$ estimated from $\mathbf{Y}$, the individual estimates of $\mathbf{G}$ (i.e., \ac{UT}s-\ac{IRS} channels) and $\mathbf{Q} = \mathbf{X} \otimes \mathbf{H}$ can be obtained by means of the \ac{KRF} approach, which consists of solving a set of rank-one matrix approximation problems. Sequentially, from $\mathbf{Q}$, which is estimated via \ac{KRF} in the previous stage, the individual estimates of $\mathbf{H}$  and $\mathbf{X}$  can be obtained using the \ac{KF}, which consists of solving a single rank-one matrix approximation problem. The KAKF receiver accomplishes joint channel and symbol estimation in closed-form from two stages as detailed in the following.
\vspace{-2ex}
\subsection{Khatri-Rao Factorization (KRF) Stage}
\label{SEC: KRF}
Considering the noisy version of (\ref{EQ:Y}), and exploiting the knowledge of the phase shifts and coding matrices $\mathbf{S}$ and $\mathbf{W}$, respectively (and thus $\boldsymbol{\Psi}$), the \ac{BS} firstly applies a linear filtering with $\boldsymbol{\Psi}^{\textrm{H}}$, yielding
\begin{equation}
\mathbf{Z} \doteq (1/K)\mathbf{Y}\boldsymbol{\Psi}^{\textrm{H}} = \mathbf{G} \diamond \mathbf{Q} + \bar{\mathbf{N}} \; \in \mathbb{C}^{ITM \times P} ,
\label{EQ:Y noise}
\end{equation}
where $\bar{\mathbf{N}} = \mathbf{N}\boldsymbol{\Psi}^{\textrm{H}}$ represents the equivalent filtered noise term. From (\ref{EQ:Y noise}), the individual estimates of $\Hat{\mathbf{G}}$ and $\Hat{\mathbf{Q}}$ can be obtained from their noisy Khatri-Rao product by solving 
\begin{equation}
   \left(\hat{\mathbf{G}},\hat{\mathbf{Q}}\right)  = \underset{\{\mathbf{G},\mathbf{Q}\}}{\min}\left\|\mathbf{Z} - \mathbf{G} \diamond \mathbf{Q}\right\|_{\text{F}}^2.
   \label{EQ: OPT_KRF}
    \end{equation}
Defining $\boldsymbol{\mu}_p$ as the $p$-th column of $\mathbf{Z}$, we have $\boldsymbol{\mu}_{p} \doteq (\mathbf{g}_{p} \otimes \mathbf{q}_{p}) + \mathbf{n}_p$, where $\mathbf{n}_p$ is the corresponding column of the noise matrix in  (\ref{EQ:Y noise}), and $\mathbf{g}_{p} \in \mathbb{C}^{I \times 1}$ and $\mathbf{q}_{p} \in \mathbb{C}^{TM \times 1}$ corresponds to the $p$-th column of the matrices $\mathbf{G}$ and $\mathbf{Q}$, respectively. Using the property $\mathbf{g}_{p} \otimes \mathbf{q}_{p} = \textrm{vec}(\mathbf{q}_{p}\mathbf{g}_{p}^{\textrm{T}})$, Problem (\ref{EQ: OPT_KRF}) can be equivalently recast as
\begin{equation}
    \left(\hat{\mathbf{G}},\hat{\mathbf{Q}}\right) = \underset{\{\mathbf{q}_{p}, \mathbf{g}_{p}\}}{\arg\min}\sum_{p = 1}^{P}\left\|\Tilde{\mathbf{Z}}_{p} - \mathbf{q}_{p}\mathbf{g}_{p}^{\textrm{T}}\right\|_\text{F}^2,
    \label{appKr}
\end{equation}
where $\Tilde{\mathbf{Z}}_{p} = \textrm{unvec}_{TM \times I}(\boldsymbol{\mu}_{p})$. Note that $\widetilde{\mathbf{Z}}_{p}$ can be approximated as a rank-one matrix given by the outer product of $\mathbf{q}_{p}$ and $\mathbf{g}_{p}$. \gil{The solution of (\ref{appKr}) is given by the best \gil{rank-one approximation } 
obtained from the dominant left and right singular vectors of $\widetilde{\mathbf{Z}}_{p}$  \cite{eckart1936approximation}}.
\gil{Therefore,} the \ac{KRF} stage of the proposed receiver thus consists of solving $P$ independent rank one-approximation problems, as summarized in \textbf{Algorithm \ref{Algo:KRF}}.
\begin{algorithm}[!t]
\IncMargin{1em}
\scriptsize{
	\DontPrintSemicolon
	\DontPrintSemicolon
	\SetKwData{Left}{left}\SetKwData{This}{this}\SetKwData{Up}{up}
	\SetKwFunction{Union}{Union}\SetKwFunction{FindCompress}{FindCompress}
	\SetKwInOut{Input}{input}\SetKwInOut{Output}{output}
	\textbf{Procedure}\\
	\Input{$\mathbf{Z}$
	}
	\Output{$\hat{\mathbf{G}}$ and $\hat{\mathbf{Q}}$ }
	\BlankLine
	\Begin{
		\For{$p = 1, \dots ,P$}{
			$\Bar{\mathbf{Z}}_{p} \longleftarrow \textrm{unvec}_{TM \times I}(\boldsymbol{\mu}_{p})$\;
			$(\mathbf{u}_1,\mathbf{\sigma}_1,\mathbf{v}_1)\longleftarrow\textrm{truncated-SVD}(\Bar{{\mathbf{Z}}}_{p})$\;
			$\hat{\mathbf{g}}_{p} \longleftarrow \sqrt{\sigma_1}\mathbf{v}_1^\ast$\,, \textrm{where $\sigma_1$ is the dominant singular value}\;
			$\hat{\mathbf{q}}_{{p}} \longleftarrow \sqrt{\sigma_1}\mathbf{u}_1$\;	
			\textbf{end}}
		\textit{Reconstruct} $\hat{\mathbf{Q}}$ \textit{and} $\hat{\mathbf{G}}$:\;
		$\hat{\mathbf{Q}} \longleftarrow \left[\hat{\mathbf{q}}_{{1}}, \dots , \hat{\mathbf{q}}_{{P}}\right]$; \quad
		$\hat{\mathbf{G}} \longleftarrow \left[\hat{\mathbf{g}}_1, \dots , \hat{\mathbf{g}}_{P}\right]$\;
		\textit{Remove the scaling ambiguities of $\hat{\mathbf{Q}}$ \textit{and} $\hat{\mathbf{G}}$.}\;
		\textbf{end}
	}
	\caption{Khatri-Rao Factorization (KRF) Stage 
	}
	\label{Algo:KRF}
	}
\end{algorithm}
\vspace{-4ex}
\subsection{Kronecker Factorization Stage}
\label{Sec:KrD algorith}
\gil{Recall that the estimate of $\hat{\mathbf{Q}}$ is delivered from the \ac{KRF} stage in  section \ref{SEC: KRF}. From (\ref{EQ:Vec of Y_k,i}) and (\ref{EQ:y_k}), we have that $\hat{\mathbf{Q}}$ can be approximated as the Kronecker product \gf{of} the transmitted data matrix and \ac{IRS}-\ac{BS} channel matrix, i.e.}
\begin{equation}
\hat{\mathbf{Q}} \approx \mathbf{X} \otimes \mathbf{H}\; \in \mathbb{C}^{TM \times P}.
\label{kronP}
\end{equation}
According to the Kronecker product definition, $\hat{\mathbf{Q}}$ can be seen as the following block matrix
\begin{equation}
	\hat{\mathbf{Q}} = \left[\begin{array}{ccc}
		\hat{\mathbf{Q}}_{1,1} &  \dots & \hat{\mathbf{Q}}_{1,UL}\\
		\vdots &  \ddots & \vdots\\
		\hat{\mathbf{Q}}_{T,1} & \dots & \hat{\mathbf{Q}}_{T,UL}\\
	\end{array}\right] \in \mathbb{C}^{TM \times P},
	\label{16}
\end{equation}
where each of the sub-matrices represents a scaled version of $\mathbf{H}$, i.e.,
\begin{equation}
\hat{\mathbf{Q}}_{t,j} = x_{t,j}\mathbf{H},
\label{17}
\end{equation}
for $t = 1, \ldots, T$ and $j = 1, \ldots, UL$. Therefore, the individual estimates for $\hat{\mathbf{X}}$ and $\hat{\mathbf{H}}$ can be obtained from $\hat{\mathbf{Q}}$ by solving
\begin{equation}
    (\hat{\mathbf{X}}, \hat{\mathbf{H}}) = \underset{\mathbf{X}, \mathbf{H}}{\min}\|\Hat{\mathbf{Q}} - \mathbf{X} \otimes \mathbf{H}\|_\text{F}.
    \label{EQ:Cost Func KrD}
\end{equation}
The matrix $\hat{\mathbf{Q}}$ can be properly rearranged by \gf{stacking} the vectorized form of the blocks $\hat{\mathbf{Q}}_{t,j}$ in (\ref{17}), as in \cite{Potsianis_Thesi}, so that a rank-one matrix $\widetilde{\mathbf{Q}}$ is constructed, as follows
\begin{equation}
	\begin{split}
 	\widetilde{\mathbf{Q}} & = \left[\begin{array}{ccc}
 		x_{1,1}\textrm{vec}(\mathbf{H}) \, &
 		\dots \,
 		&
 		x_{T,UL}\textrm{vec}(\mathbf{H})
 	\end{array}\right]^{\textrm{T}} \\
  & = \textrm{vec}(\mathbf{X}) \textrm{vec}(\mathbf{H})^{\textrm{T}} \in \mathbb{C}^{(TUL) \times (NM)}.
 \label{EQ: Q_Tilde}
   \end{split}
\end{equation}
%
Therefore, the problem in (\ref{EQ:Cost Func KrD}) becomes equivalent to solving the following rank-one matrix approximation problem
\begin{equation}
    (\hat{\mathbf{X}}, \hat{\mathbf{H}}) = \underset{\mathbf{X}, \mathbf{H}}{\min}\|\widetilde{\mathbf{Q}} - \mathbf{x}\mathbf{h}^{\textrm{T}}\|_\text{F},
    \label{EQ:Cost Func KrD _Rank_One}
\end{equation}
where $\mathbf{x} = \textrm{vec}(\mathbf{X})$ and $\mathbf{h} = \textrm{vec}(\mathbf{H})$. The best estimates to $\mathbf{x}$ and $\mathbf{h}$ (and consequently to $\mathbf{X}$ and $\mathbf{H}$) are obtained by truncating the \gf{\ac{SVD}} of $\widetilde{\mathbf{Q}}$ to its rank-one approximation. Therefore, the \ac{KF} stage of the proposed receiver is solved from a single rank-one matrix approximation step, as shown in \textbf{Algorithm \ref{Algo:Kron}}.   
\begin{algorithm}[!t]
\IncMargin{1em}
\scriptsize{
	\DontPrintSemicolon
	\DontPrintSemicolon
	\SetKwData{Left}{left}\SetKwData{This}{this}\SetKwData{Up}{up}
	\SetKwFunction{Union}{Union}\SetKwFunction{FindCompress}{FindCompress}
	\SetKwInOut{Input}{input}\SetKwInOut{Output}{output}
	\textbf{Procedure}\\
	\Input{$\hat{\mathbf{Q}}$
	}
	\Output{$\hat{\mathbf{X}}$ and $\hat{\mathbf{H}}$ }
	\BlankLine
	\Begin{
	\begin{enumerate}
	    \item[1.] Construct the rank-one matrix $\widetilde{\mathbf{Q}} \in \mathbb{C}^{TLU \times MN}$ from $\hat{\mathbf{Q}}$.
	    \item[2.] $(\mathbf{u}_1,\mathbf{\sigma}_1,\mathbf{v}_1)\longleftarrow\textrm{truncated-SVD}(\widetilde{\mathbf{Q}})$\;
	    $\hat{\mathbf{x}} \longleftarrow \sqrt{\sigma_1}\mathbf{u}_1$;\quad
	    $\hat{\mathbf{h}} \longleftarrow \sqrt{\sigma_1}\mathbf{v}_{1}^{\ast}$
	    \item[3.] Reconstruct $\hat{\mathbf{X}}$ and $\hat{\mathbf{H}}$ by unvec $\hat{\mathbf{x}}$ and $\hat{\mathbf{h}}$
	    \item[4.] Remove the scaling ambiguities of $\hat{\mathbf{X}}$ and $\hat{\mathbf{H}}$.
	\end{enumerate}
	\textbf{end}
	    }
	\caption{Kronecker Factorization Stage
	}
	\label{Algo:Kron}
	}
\end{algorithm}
\vspace{-2ex}
\subsection{Identifiability Analysis and Computational Complexity}
The necessary condition on the identifiability of the proposed KAKF receiver is linked to the linear filtering step in (\ref{EQ:Y noise}). It requires that the designed matrix $\boldsymbol{\Psi}$ to be full row-rank, implying that $K \geq P$. This condition establishes a lower-bound on the number of transmission blocks necessary for the proposed receiver to jointly estimate the channels and data symbols of all the users. This indicates that the required number of transmission blocks scales with $L$, $N$ and $U$ at least linearly, as $P = NLU$. Note that the computational complexity in both stages  (Algorithms 1 and 2) of the proposed KAKF receiver is dominated by the truncated \ac{SVD}. \gilm{This truncated \ac{SVD} can be efficiently obtained from partial QR factorization schemes \cite{Halko2011}. The total complexity of KAKF receiver is given by $\mathcal{O}(PTM) + P\mathcal{O}(TMI)$, since the \ac{KF} stage involves a single rank-$1$ approximation step, while the \ac{KRF} one has $P$ rank-$1$ approximation steps.} 
\gild{It is worth noting that in the \ac{KRF} stage the $P$ factors of the Khatri-Rao product can be estimated independently. Hence, the processing delay can be controlled by executing the $P$ estimation steps \gf{in parallel processors at the \ac{BS}.}} \gild{The pilot-assisted method \cite{Alexander_SAM2020}, used as benchmark, is an iterative solution where the computational complexity is dominated by the pseudo-inverse calculation such that the total complexity is $I_{\max}\mathcal{O}(2N^3 + 4N^2K(L+M) - NK(L+M))$.}

\vspace{-2ex}
\subsection{Scaling Ambiguities}
\label{Sec:ambigudy}

Once the rank-one approximations are computed via truncated \acp{SVD}, the estimates provided by the \ac{KRF} and \ac{KF} stages are unique up to scaling ambiguities. From the \ac{KRF} stage, the following estimates are obtained $ \Hat{\mathbf{G}} = \mathbf{G}\boldsymbol{\Delta}_{\mathbf{G}}$ and $\Hat{\mathbf{Q}} = \mathbf{Q}\boldsymbol{\Delta}_{\mathbf{Q}}$,
where $\boldsymbol{\Delta}_{\mathbf{G}}$ and $\boldsymbol{\Delta}_{\mathbf{Q}}$ are diagonal matrices that contain the column scaling ambiguities such that $\boldsymbol{\Delta}_{\mathbf{G}}\boldsymbol{\Delta}_{\mathbf{Q}} = \mathbf{I}_{P}.$ To remove these scaling ambiguities, the \ac{BS} needs to know one row of $\mathbf{G}$ or $\mathbf{Q}$. Such a scaling can be handled by assuming that the first row of $\mathbf{Q}$ is known. This is equivalent to \gil{assuming} that the \ac{BS} has the knowledge of the first row of $\mathbf{X}$ and the first row of the \ac{BS}-\ac{IRS} channel $\mathbf{H}$. In practice, the \ac{BS} can estimate the first row of $\mathbf{H}$ based on a simple scheme proposed in \cite{Hu2021}, where the \ac{IRS} reflects back pilots sent by the \ac{BS}.  

 \section{Simulation Results}
  \begin{figure*}[!t]
    \centering
    \subfloat[\vspace{-1ex} Channels NMSE versus SNR.]{\includegraphics[scale=0.4]{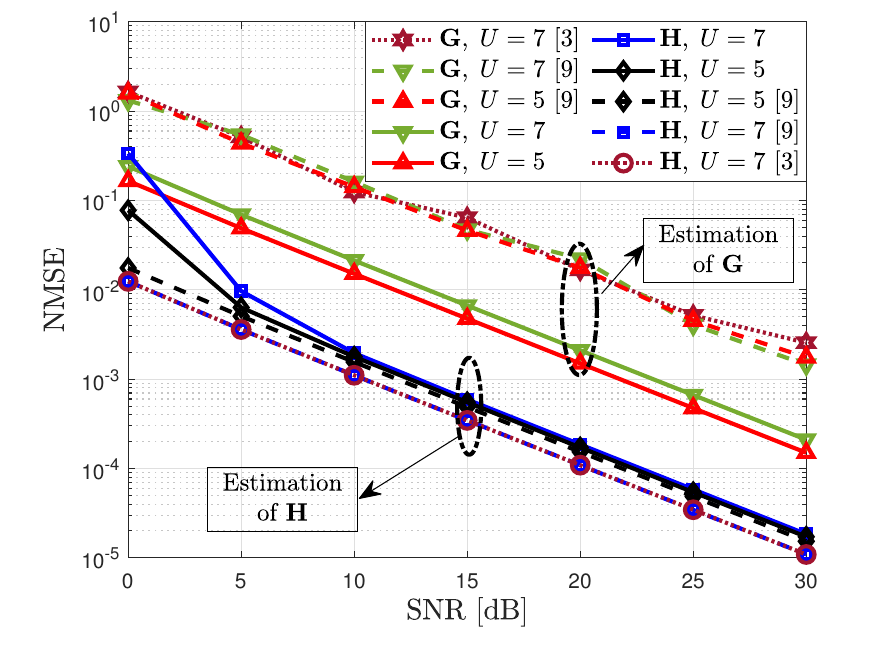}\label{Fig: NMSE}} 
    \subfloat[\vspace{-1ex} Average runtime versus SNR.]{\includegraphics[scale=0.4]{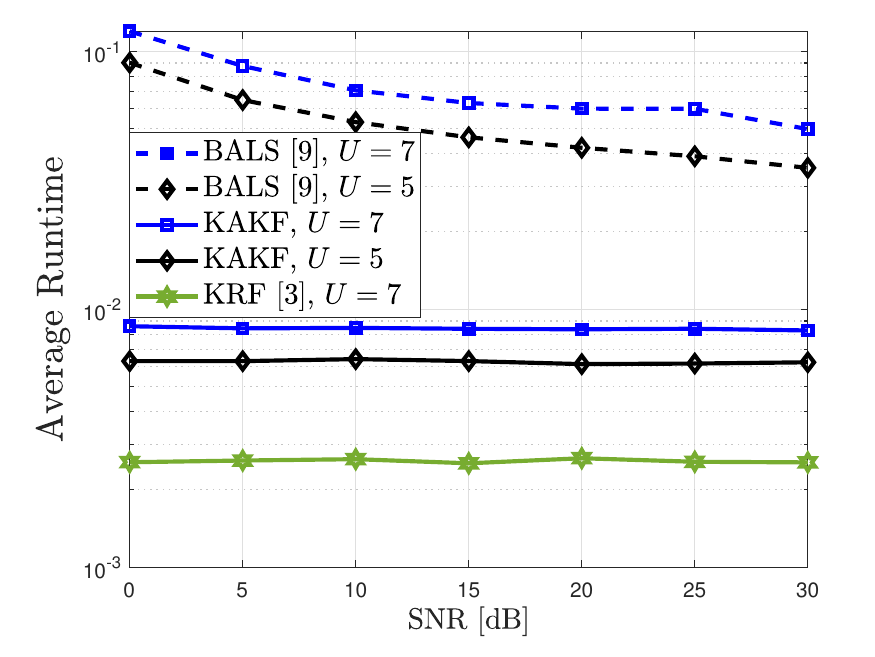}\label{Fig: complexity}}
    \subfloat[ \vspace{-1ex} SER versus SNR.]{\includegraphics[scale=0.4]{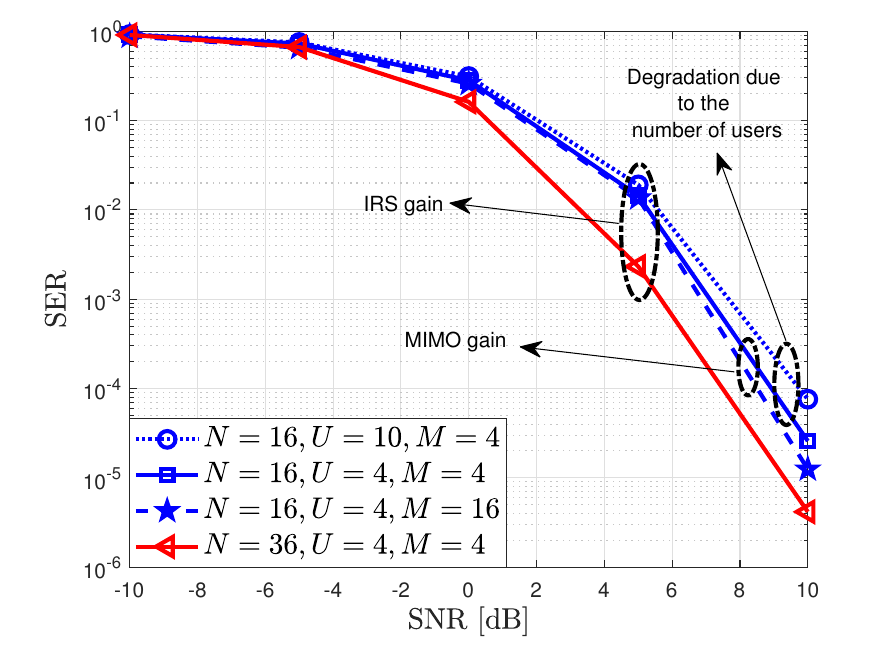} \label{Fig: SER}}
    \caption{Performance evaluation of the proposed KAKF receiver in terms of channels NMSE, average runtime and SER.}
\end{figure*}
In this section, the performance of the proposed two-stage KAKF semi-blind receiver is evaluated in terms of the \ac{NMSE} of the estimated channels, average \ac{SER}, and average runtime metrics. 
Particularly, the \ac{NMSE} of the \ac{BS}-\ac{IRS} \ac{CE} is defined as $\textrm{NMSE}(\Hat{\boldsymbol{\Omega}}) = (\sfrac{1}{R})\sum_{r=1}^{R} \sfrac{\|\boldsymbol{\Omega}_{(r)} - \Hat{\boldsymbol{\Omega}}_{(r)}\|_\text{F}^2}{ \|\boldsymbol{\Omega}_{(r)}\|_\text{F}^2}$, where $\boldsymbol{\Omega}$ = $\mathbf{H}$ or $\mathbf{G}$, and $\Hat{\boldsymbol{\Omega}}_{(r)}$ represents the estimation of the channel at the $r$-th Monte Carlo run. The results are averaged over $R = 10^{4}$ independent Monte Carlo runs. We assume that the involved channels follow geometric models such that $\mathbf{H} = \mathbf{A}_{\textrm{IRS}}\textrm{diag}(\boldsymbol{\beta})\mathbf{A}_{\textrm{BS}}^{\textrm{H}}$ and $\mathbf{G} = \mathbf{B}_{\textrm{UT}}\textrm{diag}(\boldsymbol{\gamma})\mathbf{B}_{\textrm{IRS}}^{\textrm{H}}$ (for each \ac{UT}), in which the number of dominant paths in the \ac{IRS}-\ac{BS} and \ac{UT}s-\ac{IRS} links are denoted by $L_h$ and $L_g$\footnote{For simplicity, in the simulations, we assume that each \ac{UT} contributes with the same number $L_{g}$ of dominant paths.}, respectively, while $\boldsymbol{\beta}$ and $\boldsymbol{\gamma}$ denote the channel gain vectors. The departure and arrival angles are randomly and uniformly distributed between $0$ and $2\pi$, while the path gains follow a complex Gaussian distribution with zero mean and unitary variance. The transmitted data symbols are chosen from a $16$-PSK alphabet, and each UT has a transmission rate of \gild{$\rho = \sfrac{[(T - 1) + T_D]\log_2(M)L}{T_c}$ for a given coherence time $T_c$ and data transmission time~$T_D$.}%

Figure \ref{Fig: NMSE} depicts the \ac{NMSE} performance of the proposed semi-blind KAKF receiver as a function of the signal-to-noise ratio (SNR). \gild{As a performance benchmark, we also depict the results of the  \ac{BALS} method proposed in \cite{Alexander_SAM2020}, as well as the results of the closed-form \ac{KRF} solution proposed in \cite{Gil_JTSP}, which are pilot-assisted methods.} Although the competitive methods \gf{do not} consider \ac{UT}s-\ac{IRS} time-varying channels, we adapt \gf{them} to the time-varying case to ensure a fair comparison. In this experiment, we assume $N = 36$, $M= 4$, $U \in \{5, 7 \}$, each \ac{UT} equipped with $2$ antennas ($L = 2$). Besides, we consider $L_{h} = 1$ and $L_{g} = 1$. Further, the \ac{BS} captures $I = 5$ \ac{UT}s-\ac{IRS} channel variations and $\{T, K\} = \{2, 720\}$. As it can be observed, our proposed semi-blind KAKF receiver \gf{significantly outperforms the \ac{BALS} method} in the accuracy of the estimation of the \ac{UT}s-\ac{IRS} channel $\mathbf{G}$. As for the \ac{IRS}-\ac{BS} channel $\mathbf{H}$, the \ac{BALS} method outperforms the proposed receiver in the low \ac{SNR} regime for $U = 5$. However, in the medium-to-high \ac{SNR} regime, the two receivers \gf{show} similar performances. Besides, the \ac{BALS} estimator, which requires pilot sequences, has a very modest gain for $U = 7$. 
This is an expected result since \gild{the \ac{IRS}-\ac{BS} channel $\mathbf{H}$ is estimated in the second \ac{KF} stage \gf{when using the KAKF receiver}. Therefore, it is affected by the error propagation originating from the first \ac{KRF} stage. This explains its worse performance in estimating $\mathbf{H}$ in the low \ac{SNR} regime} compared to \ac{BALS}, in which each channel matrix is estimated in an alternating way, thus avoiding error propagation. \gild{Note that the \ac{KRF} \cite{Gil_JTSP} performance is similar to the \ac{BALS} performance.} However, considering the end-to-end channel, \gil{our} proposed scheme offers a remarkable improvement in the \ac{CE} accuracy. For an \ac{SNR} equal to 20 dB, while KAKF and \ac{BALS} have similar \ac{NMSE} performances for the estimation of $\mathbf{H}$, the proposed scheme offers an order of magnitude improvement in the estimation of $\mathbf{G}$ (see Fig.~\ref{Fig: NMSE}).

To evaluate the computational complexity of the proposed semi-blind KAKF receiver, we consider the same parameters as in Fig. \ref{Fig: NMSE} and plot in Fig. \ref{Fig: complexity} the average runtime as a function of the \ac{SNR}. Since the KAKF receiver has a closed-form solution, it offers a considerable complexity reduction compared to the iterative \ac{BALS} receiver \cite{Alexander_SAM2020}. \gild{On the other hand, the KRF method \cite{Gil_JTSP} is a bit faster than our proposed method, since the KAKF receiver involves an additional processing step to estimate the symbol matrix.} Furthermore, the computational complexity of KAKF does not depend on the SNR, unlike \ac{BALS}, where the number of iterations for convergence increases for lower SNR values. This is an interesting result, since the \ac{CE} performance and the computational complexity depend on the number of \gf{parameters to be estimated, which} scales with the number of \ac{UT}s. In particular, the estimate of $\mathbf{H}$ is more accurate than that of $\mathbf{G}$, since the latter has more coefficients to be estimated. However, even in this situation, KAKF has a much lower runtime compared to the method of \cite{Alexander_SAM2020} and, thus, an improved end-to-end latency, \gild{for the same data block length}.

The \ac{SER} performance is illustrated in Fig. \ref{Fig: SER}\footnote{\color{black}In practice, in the \ac{BALS} and \ac{KRF} methods, optimizing the beamforming weights for data transmission takes place prior to data transmission. However, this optimization step is out of the scope of the present paper.}
considering different number\gil{s} of users $U \in \{4, 10\}$, different number\gil{s} of antennas $M \in \{4, 16\}$, and different number\gil{s} of \ac{IRS} elements $N \in \{16, 36\}$. Besides, $K = NLU$ and $T = 4$. The other parameters are the same as in Fig. \ref{Fig: NMSE}. This result shows that the \ac{SER} \gf{decreases} as a function of the number of \ac{IRS} elements. This is an interesting result since no optimization process is carried out, which may further improve the system performance. Fixing $N= 16$, the effect of the number of antennas at the \ac{BS} and the number of users is shown. By increasing $M$ and keeping $U = 4$, the \ac{SER} improves due to a spatial diversity gain. Changing the roles, i.e., keeping $M= 4$ and increasing $U$, the \ac{SER} degrades as $U$ increases. This \gil{is intuitive}, since the number of channel components to be estimated increases as more users are active in the system. 

\vspace{-3.5ex}
\section{Conclusion}
\gilm{We proposed a two-stage closed-form semi-blind receiver for joint channel and symbol estimation in \ac{IRS}-assisted MU-MIMO systems based on a generalized PARATUCK tensor modeling. The so-called KAKF receiver provides joint estimates of the \ac{UT}s-\ac{IRS} and \ac{IRS}-\ac{BS} channels and the transmitted symbols with low computational complexity. Compared to its pilot-assisted competitor, KAKF yields more accurate channel estimates while handling time-varying channels.}

\vspace{-1ex}

\renewcommand\baselinestretch{.95}



\end{document}